%% file: main.tex
\documentclass[conference]{IEEEtran}
\IEEEoverridecommandlockouts
\usepackage{cite}
\usepackage{amsmath,amssymb,amsfonts}
\usepackage{algorithmic}
\usepackage{graphicx}
\usepackage{textcomp}
\usepackage{xcolor}
\usepackage{subfig}
\usepackage[font=small]{caption}

\def\BibTeX{{\rm B\kern-.05em{\sc i\kern-.025em b}\kern-.08em
    T\kern-.1667em\lower.7ex\hbox{E}\kern-.125emX}}
\begin{document}

\title{CAIM: Cooperative Angle of Arrival Estimation using the Ising Method\\
}

\author{\IEEEauthorblockN{Shiva Akbari}
\IEEEauthorblockA{\textit{Dept. of Electrical and Computer Engineering} \\
\textit{University of Toronto}\\
Toronto, Ontario Canada \\
shiva.akbari@mail.utoronto.ca}
\and
\IEEEauthorblockN{Shahrokh Valaee}
\IEEEauthorblockA{\textit{Dept. of Electrical and Computer Engineering} \\
\textit{University of Toronto}\\
Toronto, Ontario, Canada \\
valaee@ece.utoronto.ca}
}

\maketitle

\begin{abstract}
This paper proposes a cooperative angle-of-arrival (AoA) estimation, taking advantage of co-processing channel state information (CSI) from a group of access points that receive signals of the same source. Since received signals are sparse, we use Compressive Sensing (CS) to address the AoA estimation problem. We formulate this problem as a penalized $\ell_0$-norm minimization, reformulate it as an Ising energy problem, and solve it using Markov Chain Monte Carlo (MCMC). Simulation results show that our proposed method outperforms the existing methods in literature.
\end{abstract}


\input{Introduction}
\input{Background}

\input{Problem_formulation}
\input{Performance_evaluation}
\input{Conclusion}
\section*{Acknowledgment}
The authors would like to thank Fujitsu Laboratories Ltd. and Fujitsu Consulting (Canada) Inc. for providing financial support and access to Digital Annealer at the University of Toronto.

\bibliographystyle{IEEEtran}

\end{document}

%% file: Introduction.tex
\section{Introduction}
\label{sec.introduction}
In recent years, location-based services are getting more popular, and accurate localization is becoming crucial. 
WiFi-based localization methods are widely used in positioning systems and are able to provide accurate location services especially in indoor environments \cite{Liu2019}.
WiFi-based localization methods can be categorized into geometry-based methods, and fingerprint-based methods.  Geometry-based methods widely utilize the angle 
of arriving wavefront,
called the angle-of-arrival (AoA), for target localization \cite{Liu2019, Xiong2013,kotaru2015,Rahman2018, Rahman2019, Gong2019}. 


Some recent approaches for AoA estimation use the theory of Compressive Sensing (CS) 
\cite{Feng2012, Bazzi2016, garcia2017, Gong2019, Han2020}. These approaches take advantage of the signal sparsity, and compared to other techniques, use fewer observations to estimate AoA. Moreover, the CS localizer  does not need to know the number of sources.
In some CS-based techniques, such as \cite{Gong2019} and \cite{Malioutov2005}, the AoA estimation problem is formulated as a convex $\ell_1$-norm optimization problem. However, this relaxation can impose a bias on the estimation \cite{Northardt2013}. 
In \cite{Gong2019}, after solving the $\ell_1$-norm optimization problem, the results are combined by forming a weighted mean squared error optimization problem. This problem is then solved using convex optimization methods. In \cite{Han2020}, a regularized $\ell_0$-norm formulation for the AoA estimation problem for one access point is presented to overcome the inherent bias due to the $\ell_1$-norm relaxation.

\textcolor{black}
{In this paper, we propose a cooperative AoA estimation method for multiple access points. In order to co-process the received {\it channel state information} (CSI) from all the access points, we take advantage of the fact that the received AoA of two access points from a target in far-field will be the same if we rotate one of the access points to have the same orientation as the other one.} 

\begin{figure}
\centering
\subfloat[Before rotation]{\includegraphics[width = 1.1in]{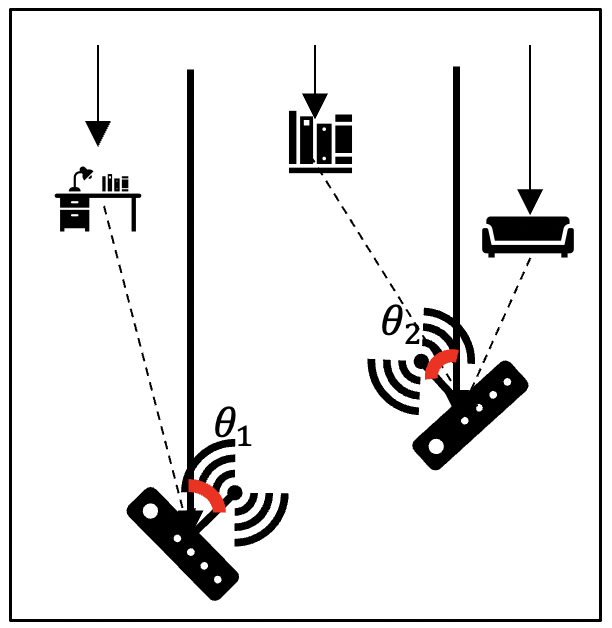}} ~~~~~~
\subfloat[After rotation with angle $\alpha_{(1,2)}$ ]{\includegraphics[width = 1.1in]{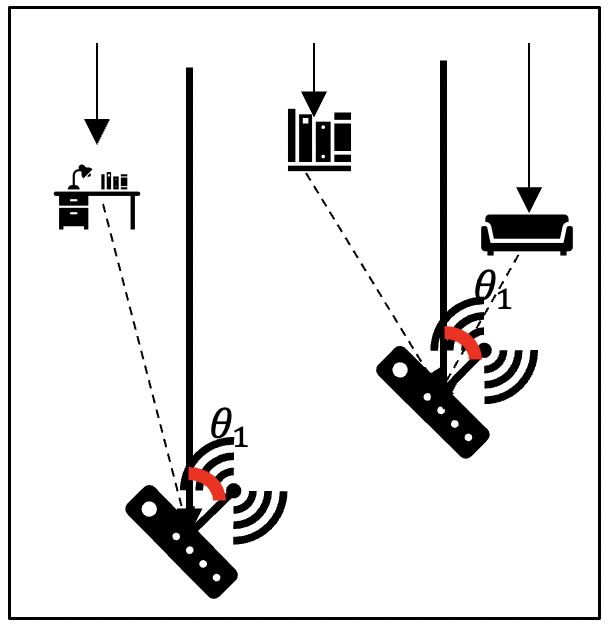}}\\
\caption{A simple scenario with 2 access points}
\label{fig:fig0}
\end{figure}

To clarify this observation, we give an example of a simple scenario. As shown in Fig. \ref{fig:fig0}, we have 2 access points, a source and some reflectors inside the environment. Each access point receives signals both from the source and the reflectors. The source is located at the far-field and the signals arrive at the access points as a planar wavefront. To be more specific, in the scenario specified in Fig. \ref{fig:fig0}, access point 1 receives a signal from the source with angle $\theta_1$ and another signal from a reflector. Similarly, access point 2 receives a signal from the source with angle $\theta_2$, along with signals from two reflectors. Because the source is in far-field, if we rotate access point 2 with the angle $\alpha_{(1,2)} = \phi_1 - \phi_2$, then the received AoA of LoS paths of both access points will align with each other. In other words, if the received AoA of access point 2 after rotation is  $\hat{\theta_2} = \theta_2 - \alpha_{(1,2)}$, then $\hat{\theta_2}$ and $\theta_1$ will be the same. However, because reflections are coming from different points in space, it is less likely for the reflected paths to align. Thus, by co-processing the aligned received signals, we can increase the resolution of  AoA estimator.


Motivated by \cite{Han2020}, we formulate the AoA estimation as a regularized $\ell_0$-norm minimization problem. 
This NP-hard optimization problem is transformed to an Ising energy problem and is solved using the Markov Chain Monte Carlo (MCMC) method \cite{aramon2018}.
We call our approach {\it Cooperative AoA estimation via  Ising Model} (CAIM).
Simulation results show that CAIM outperforms  both AIM \cite{Han2020}, and RoArray \cite{Gong2019}
by a significant margin.


The rest of this paper is organized as follows. In Section \ref{sec.background}, the AoA estimation problem is formulated as an optimization problem using CS. In Section \ref{sec.problem_fromulation}, the cooperative AoA estimation problem is formulated and its transformation to an Ising energy problem is explained. Section \ref{sec.performance_evaluation} presents the simulation results. Finally, Section \ref{sec.conclusion} concludes the paper.

%% file: Background.tex
\section{Basis Pursuit Formulation}
\label{sec.background}

Consider $P$ access points each equipped with a uniform linear array (ULA) antenna with $M$ elements. The orientations of antenna arrays with respect to a global coordinate system are known and given by $\phi_1, \phi_2, \cdots, \phi_P$.
 A far-field wave arriving at the $p$th array from the angle $\theta_p$ made with the broadside of the array creates the vector
\begin{equation}
\label{back.eq1}
    \textbf{y}_p = \textbf{a}(\theta_p)s_p + \textbf{n}_p
\end{equation}
where $\textbf{y}_p$ is the $M \times 1$ vector of received signal, $s_p$ is the complex signal amplitude, $\textbf{n}_p$ is the $M \times 1$ noise vector, which is assumed to be Gaussian and independent of the signal, and $\textbf{a}(\theta_p)$ is the $M \times 1$ array manifold vector. For a ULA, the array manifold is given by
\begin{equation}
\label{back.eq2}
    \textbf{a}(\theta_p) = \begin{bmatrix} 1, e^{-j2\pi \frac{d}{\lambda}sin(\theta_p)},\cdots,e^{-j2\pi \frac{d}{\lambda}(M-1)sin(\theta_p)} \end{bmatrix} ^T
\end{equation}
where $d$ is the distance between two adjacent array elements and $\lambda$ is the wavelength. If there are multiple signals arriving from $N$ distinct angles, the received signal will be
\begin{equation}
\label{back.eq3}
    \textbf{y}_p = \textbf{A}_p(\textbf{$\theta$}_p)\textbf{s}'_p + \textbf{n}_p
\end{equation}
where $\textbf{A}_p(\textbf{$\theta$}_p)$ is the $M \times N$ array manifold matrix, $\textbf{$\theta$}_p$ is the  vector of AoAs of the signals, and $\textbf{s}'_p$ is the $N \times 1$ vector of signal components.
An AoA estimator finds the best $\textbf{$\theta$}_p$ that satisfies the above equation given the observation vector $\textbf{y}_p$. 

An effective method to estimate $\textbf{$\theta$}_p$ is based on \textit{Compressive Sensing} (CS).
In the CS theory, 
an $N_r \times 1$ sparse signal 
can be reconstructed with M measurements, where $ M \ll N_r$ \cite{Candes2008}.
Inspired by this observation,  we build an extended array manifold matrix by digitizing the angular search space into a grid of size $N_r$, and selecting $\textbf{$\Psi$}$ such that
\begin{equation}
\label{back.eq4}
    \textbf{y}_p = \textbf{$\Psi$} \textbf{s}_p + \textbf{n}_p
\end{equation}
where $\textbf{$\Psi$}$ is the $M \times N_r$ over-complete array manifold matrix, and $\textbf{s}_p$ is the $N_r \times 1$ sparse vector of signal components  with a support set of size $M \ll N_r$. An estimate of the AoA is obtained by forming the basis pursuit (BP) problem
\begin{equation}
\label{back.eq5}
    \begin{aligned}
        \min_{\textbf{s}_p} \quad &\vert \vert\textbf{s}_p \vert \vert_0\\
        \textrm{s.t.} \quad &\vert \vert \textbf{y}_p- \textbf{$\Psi$}\textbf{s}_p \vert \vert_2 ^2< \epsilon
    \end{aligned}
\end{equation}
The BP problem is then solved for all the $P$ access point antenna arrays to find the corresponding AoAs \cite{Han2020}.

In this paper, we seek methods that combine the signal measurements of different access points such that they can be processed jointly to detect the true AoA. The co-processing of the signals from all arrays gives a higher resolution compared to separate processing of individual arrays.


%% file: Problem_formulation.tex
\section{Joint  Estimation of AoA}
\label{sec.problem_fromulation}

We observe that if a rotation of angle $\alpha_{(p,q)} = \phi_p - \phi_q$, 
is applied to the $q$th access point, then the signal arriving from the line-of-sight (LoS)  for access point $p$ and access point $q$ will 
have the same index in the sparse signal vectors. Let us denote the rotated version of $\textbf{s}_q$ by $\textbf{s}_q ^{(\alpha_{(p,q)})}$. Then both $\textbf{s}_p$ and $\textbf{s}_q ^{(\alpha_{(p,q)})}$ will have non-zero values at the index corresponding to  LoS. In other words, the support sets of $\textbf{s}_p$ and $\textbf{s}_q ^{(\alpha_{(p,q)})}$ will overlap at an index that corresponds to LoS. Fortunately, the reflections are less likely to have this property due to the fact that reflections for different access points do not originate from the same location in space and have a random nature.


Based on the above observation, we propose the following optimization problem,
\begin{equation}
    \label{pf.eq1}            \min_{\textbf{S}}\quad \sum_{p = 1}^{P} \big(\vert \vert\textbf{s}_p \vert \vert_0 + \gamma
        \vert \vert \textbf{y}_p- \textbf{$\Psi$}\textbf{s}_p \vert \vert_2^2\big)  + \mu g(\textbf{S}, \textbf{S}^{(\alpha)})
\end{equation}
where $g(\textbf{S}, \textbf{S}^{(\alpha)})$ is a function that penalizes the objective function when the nonzero elements of its arguments are not aligned, and $\gamma$ and $\mu$ are the regularization parameters. 
In (\ref{pf.eq1}), $\textbf{S}$ is the set of $\{\textbf{s}_p$, $p=1, \cdots, P\}$ and $\textbf{S}^{(\alpha)}$ is the set of rotated version of $\textbf{S}$,  i.e.  $\{\textbf{s}_q ^{(\alpha_{(p,q)})}, p= 1,\cdots, P-1, q = p+1, \cdots, P\}$.

Let us define the $N_r \times 1$ vector $\textbf{x}_p$ as
\begin{equation}
\label{pf.eq2}
    \textbf{x}_p = \begin{bmatrix} I(s_{p,1}), I(s_{p,2}),\cdots, I(s_{p, N_r}) \end{bmatrix}^ {T}, p = 1, 2, \cdots P
\end{equation}
where $s_{p,i}$ is the $i$th element of $\textbf{s}_p$, and $I(.)$ is the identifier function, which is 1 if its argument is nonzero and 0 otherwise. Similarly, we define the binary variables  $\textbf{x}_p^{(\alpha_{(p,q)})}$, for $p = 1, 2, \cdots, P-1$, and $q = p+1,\cdots,P$. The function $g(\textbf{S}, \textbf{S}^{(\alpha)})$ takes the following form
\begin{equation}
\begin{aligned}
\label{pf.eq3}
    g(\textbf{S},\textbf{S}^\alpha) = \sum_{p=1}^{P-1}\sum_{q=p+1}^{P} \sum_{i=1}^{N_r} (x_{p,i} + x_{q, i}^{(\alpha_{(p,q)})} - 2x_{p,i}x_{q,i}^{(\alpha_{(p,q)})})
    \end{aligned}
\end{equation}
where $x_{p,i}$ is the $i$th element of $\textbf{x}_p$. 
The function $g(\textbf{S}, \textbf{S}^{(\alpha)})$  penalizes the objective function when the elements of $\textbf{S}$ and $\textbf{S}^{(\alpha)}$ are non-zero in the same index. In other words, $g(\textbf{S}, \textbf{S}^{(\alpha)})$ reflects a pair-wise comparison of $\textbf{S}$ and $\textbf{S}^{(\alpha)}$. There are different criteria to measure the similarity of two vectors. Hamming distance, Euclidean distance and $\ell_{1}$-norm distance are among the most important ones. For our case, in which the vectors are binary, all of these three metrics translate to the XOR of the two vectors. The mathematical description of XOR is defined in (\ref{pf.eq3}). 

In the optimization problem (\ref{back.eq5}), if we restrict our variable search space to binary vectors, we will reach the following optimization problem:
\begin{equation}
\label{pf.eq4}
    \begin{aligned}
        \min_{\textbf{x}_p} \quad &\vert \vert\textbf{x}_p \vert \vert_0\\
        \textrm{s.t.} \quad &\vert \vert \textbf{y}_p- \textbf{$\Psi$}\textbf{x}_p \vert \vert_2 ^2< \epsilon
    \end{aligned}
\end{equation}
We observe that $\vert\vert \textbf{s}_p\vert\vert_0 = \vert\vert \textbf{x}_p\vert\vert_0, p = 1, 2, \cdots, P$. So, the objective function of the optimization problems (\ref{back.eq5}) and (\ref{pf.eq4}) are equivalent. However, due to the binary vector restriction, the feasibility space of problem (\ref{pf.eq4}) is a subset of the feasibility space of problem (\ref{back.eq5}). Thus, the solution of problem (\ref{pf.eq4}) is an upper-bound for the solution of problem (\ref{back.eq5}). Considering this discussion, we can modify the regularized minimization problem (\ref{pf.eq1}) as follows: 
\begin{equation}
    \begin{alignedat}{2}
    \label{pf.eq5}
            &\min_{\substack{\textbf{x}_p, \\p = 1, 2,\cdots,P}}\quad \sum_{p = 1}^{P} \Big( \vert \vert\textbf{x}_p \vert \vert_0 + \gamma
        \vert \vert \textbf{y}_p- \textbf{$\Psi$}\textbf{x}_p \vert \vert_2^2 \Big) \\&\qquad  + \mu\sum_{p=1}^{P-1}\sum_{q=p+1}^{P} \sum_{i=1}^{N_r} (x_{p,i} + x_{q, i}^{(\alpha_{(p,q)})} - 2x_{p,i}x_{q,i}^{(\alpha_{(p,q)})})
    \end{alignedat}
\end{equation}
Solving the optimization problem (\ref{pf.eq5}) 
gives us an approximation of the original problem (\ref{pf.eq1}).
Using a method similar to \cite{Han2020}, the AoA estimation problem, (\ref{back.eq3}), can be formulated as an Ising energy problem as will be discussed next.

\subsection{Ising Energy Modelling}
 The Ising model is a mathematical description of ferromagnetism in statistical physics \cite{glauber1963}. The variables in this model describe the spin of atoms, which are binary (either $-1$ or $+1$). 
By converting $\{ -1, +1\}$ states to $\{ 0, 1\}$, we can formulate a quadratic unconstrained binary optimization (QUBO) problem \cite{harju2012}. A  QUBO problem has the following form:
\begin{equation}
\label{pf.eq6}
    E(\textbf{x}; \textbf{b}, \textbf{W}) = -\sum_{i=1}^{K} b_i x_i - \sum_{i=1}^{K}\sum_{j=1}^{K}W_{i,j}x_ix_j
\end{equation}
where $\textbf{x}$ is a binary vector of length $K$, and $b_i$ and $W_{i,j}$ are the bias term and the connection weight, respectively.

Let us define 
\begin{equation*}
   \hat{\textbf{x}} = \begin{bmatrix} x_{1,1},\cdots,x_{1, N_r},x_{2,1},\cdots,x_{2,N_r},\cdots,x_{P,1},\cdots,x_{P,N_r}\end{bmatrix}^T 
\end{equation*}
We want to transform the objective function of problem (\ref{pf.eq5}) into an Ising Model. We do this in the following steps. Since $\hat{x}_{i} \in \{0,1\}$:
\begin{equation}
\label{pf.eq7}
  \sum_{p=1}^{P} \vert\vert  x_p \vert\vert_{0} = \sum_{i=1}^{PN_r}\hat{x}_{i}
\end{equation}
Further,
\begin{equation}
\begin{aligned}
\label{pf.eq8}
     &\sum_{p=1}^{P} \vert\vert y_p - \Psi x_{p}\vert\vert_{2}^{2} =\sum_{p=1}^{P}(y_p -\Psi x_{p})^{H} (y_p -\Psi x_{p})\\&\qquad = \sum_{p=1}^{P} y_{p}^{H}y_{p} + \sum_{p=1}^{P} x_{p}^{H} \Psi^{H}\Psi x_{p}-2\Re\{x_{p}^{H}\Psi^{H}y_{p}\}
\end{aligned}
\end{equation}
In the right-hand side of (\ref{pf.eq8}), the first term does not depend on $x$ and can be ignored. To rewrite the rest in terms of $\hat{\textbf{x}}$, we consider $\hat{x}_{i}^{2} = \hat{x}_{i}$ and rewrite the second term as:
\begin{equation}
\begin{aligned}
    &\sum_{p=1}^{P} x_{p}^{H} \Psi^{H}\Psi x_{p} =  \sum_{p=1}^{P}\sum_{m=1}^{M}\sum_{i=1}^{N_r} \vert\Psi_{m,i}\vert^{2}\hat{x}_{(p-1)N_r+i}\\\qquad& + \sum_{p=1}^{P}\sum_{m=1}^{M} \sum_{i=1}^{N_r}\sum_{j=1, j\neq i}^{N_r}\Psi_{m,i}^{*} \Psi_{m,j} \hat{x}_{(p-1)N_r+i}\hat{x}_{(p-1)N_r+j}
\end{aligned}
\end{equation}
and the last term will be 
\begin{equation}
    \sum_{p=1}^{P}2\Re\{x_{p}^{H}\Psi^{H}y_{p}\} = \sum_{i=1}^{N_r}\sum_{p=1}^{P}\sum_{m=1}^{M} 2\Re\{\Psi_{m,i}^{*}y_{p_{m}}\}\hat{x}_{(p-1)N_r+i}
\end{equation}

To rewrite the penalty term in (\ref{pf.eq5}) as an Ising model using $\hat{\textbf{x}}$, we should first 
implement $\hat{\textbf{x}}^{(\alpha)}$. We define the rotation as a circular shift in our vector. To be more specific, if $\alpha_{(p,q)} > 0$, then we should rearrange $x_p$ by moving the last $n_{\alpha_{(p,q)}}$ entries of $x_p$ to the first $n_{\alpha_{(p,q)}}$ positions while shifting the rest of entries. Note that $n_{\alpha_{(p,q)}} = \lfloor{\frac{\alpha_{(p,q)}}{N_r}}\rfloor$, with $\lfloor.\rfloor$ being the floor function. If $\alpha_{(p,q)} < 0$, we perform the reverse operation to implement $x_{p}^{(\alpha_{(p,q)})}$.
  
  The summation of the elements of a vector and its circular shifted version are the same. Therefore,
 \begin{equation}
     \sum_{p=1}^{P-1}\sum_{q=p+1}^{P} \sum_{i=1}^{N_r} (x_{p,i} + x_{q, i}^{(\alpha_{(p,q)})}) = (P-1) \sum_{i=1}^{PN_r}\hat{x}_i
 \end{equation}
For $\alpha_{(p,q)} > 0$,
\begin{equation}
\label{pf.eq10}
\begin{aligned}
    &\sum_{p=1}^{P-1}\sum_{q=p+1}^{P}\sum_{i=1}^{N_r} 2x_{p,i}x_{q,i}^{(\alpha_{(p,q)})} = \\&\qquad \sum_{p=1}^{P-1}\sum_{q=p+1}^{P}\sum_{h=1}^{n_{\alpha_{(p,q)}}} 2\hat{x}_{(p-1)N_r+h}\hat{x}_{qN_r-n_{\alpha_{(p,q)}} +h}+\\&\qquad \sum_{p=1}^{P-1}\sum_{q=p+1}^{P}\sum_{h=n_{\alpha_{(p,q)}}+1}^{N_r} 2\hat{x}_{(p-1)N_r+h}\hat{x}_{(q-1)N_r-n_{\alpha_{(p,q)}} +h}
    \end{aligned}
\end{equation}
where in the right-hand side of (\ref{pf.eq10}), the first term includes the first $n_{\alpha_{(p,q)}}$ elements of the $x_{q}^{(\alpha_{p,q})}$ in terms of $\hat{\textbf{x}}$. Similarly, for $\alpha_{(p,q)} < 0$, we have
\begin{equation}
\label{pf.eq11}
\begin{aligned}
    &\sum_{p=1}^{P-1}\sum_{q=p+1}^{P}\sum_{i=1}^{N_r} 2x_{p,i}x_{q,i}^{(\alpha_{(p,q)})} = \\&\qquad \sum_{p=1}^{P-1}\sum_{q=p+1}^{P}\sum_{h=1}^{Nr -n_{\alpha_{(p,q)}}} 2\hat{x}_{(p-1)N_r+h}\hat{x}_{(q-1)N_r+n_{\alpha_{(p,q)}} +h}+\\&\qquad \sum_{p=1}^{P-1}\sum_{q=p+1}^{P}\sum_{\substack{ h=\\N_r-n_{\alpha_{(p,q)}}+1}}^{N_r} 2\hat{x}_{(p-1)N_r+h}\hat{x}_{(q-2)N_r+n_{\alpha_{(p,q)}} +h}
\end{aligned}
\end{equation}

Considering (\ref{pf.eq7}) to (\ref{pf.eq11}), we can rewrite problem (\ref{pf.eq5}) in an Ising model format. In this case,
\begin{equation}
\begin{aligned}
\label{pf.eq12}
    &b_i = -\Big(1 + \gamma \sum_{m = 1}^{M} \vert \Psi_{m,h}\vert^2 - \gamma\sum_{m=1}^{M} 2\Re\{\Psi_{m,h}^{*}y_{p_{m}}\}\\& \qquad + (P-1)\mu\Big) \qquad i=(p-1)N_r+h,\\&\qquad\qquad\qquad\qquad\quad h= 1, 2,\cdots,N_r , p = 1, 2,...P
\end{aligned}
\end{equation}
and
\begin{equation}
\begin{aligned}
\label{pf.eq13}
W_{i,j} = \left \{ \begin{matrix} 0 & i=j \\-\sum_{m=1}^{M}\gamma \Psi_{m,i}^{*}\Psi_{m,j} & i,j \in \mathcal{L}, i\neq j \\ \mu & i,j \in \mathcal{S}
\end{matrix} \right.
\end{aligned}
\end{equation}\\
where $\mathcal{L} =\{i| i= (p-1)N_r+1,\cdots, pN_r, p = 1, 2,\cdots, P\}$ and $\mathcal{S} = \{ \mathcal{S}_1\cup \mathcal{S}_2\}$ in which
\begin{equation*}
        \mathcal{S}_1 = \{ i, j| i = (p-1)N_r+h , 1\leq h \leq N_r, j\in \hat{\mathcal{S}}\}
\end{equation*}
and
\begin{equation*}
        \mathcal{S}_2 = \{ i, j| j = (p-1)N_r+h , 1\leq h \leq N_r, i\in \hat{\mathcal{S}}\}
\end{equation*}
In other words, $\textbf{W}$ is a symmetric matrix. In addition, $\hat{\mathcal{S}} = \{\hat{\mathcal{S}}_1\cup\hat{\mathcal{S}}_2\cup\hat{\mathcal{S}}_3\cup\hat{\mathcal{S}}_4\}$ where $\hat{\mathcal{S}}_1$ and $\hat{\mathcal{S}}_2$ represent the two intervals made due to the circular shift  for $\alpha_{(p,q)} > 0$   and $\hat{\mathcal{S}}_3$ and $\hat{\mathcal{S}}_4$ describe those intervals for $\alpha_{(p,q)} < 0$. Specifically,
\begin{equation*}
\begin{aligned}
    \hat{\mathcal{S}}_1 = \{j| j &= qN_r-n_{\alpha_{(p,q)}}+h, h = 1,2,\cdots, n_{\alpha_{(p,q)}}, \\&p = 1,2,\cdots, P-1, q = p+1,\cdots,P,\alpha_{(p,q)}>0 \}
    \end{aligned}
\end{equation*}
\begin{equation*}
\begin{aligned}
    \hat{\mathcal{S}}_2 = \{&j| j= (q-1)N_r-n_{\alpha_{(p,q)}}+h, h = n_{\alpha_{(p,q)}} +1, \cdots, \\&N_r, p = 1,2,\cdots, P-1, q = p+1,\cdots,P,\alpha_{(p,q)}>0 \}
    \end{aligned}
\end{equation*}
\begin{equation*}
\begin{aligned}
    \hat{\mathcal{S}}_3 = \{j| j &= (q-1)N_r+n_{\alpha_{(p,q)}}+h, h = 1,\cdots,N_r-n_{\alpha_{(p,q)}},\\&p = 1,2,\cdots, P-1, q = p+1,\cdots,P,\alpha_{(p,q)}<0 \}
    \end{aligned}
\end{equation*}
\begin{equation*}
\begin{aligned}
    \hat{\mathcal{S}}_4 = \{j| j &= (q-2)N_r+n_{\alpha_{(p,q)}}+h, h = N_r-n_{\alpha_{(p,q)}}+1, \\&\cdots, N_r, p = 1,2,\cdots, P-1, q = p+1,\cdots,P,\\&\alpha_{(p,q)}<0 \}.
    \end{aligned}
\end{equation*}

Therefore, the optimization problem (\ref{pf.eq5}) can be rewritten as an Ising energy problem.
In this paper, we 
use \textit{Digital Annealer} (DA) 
\cite{aramon2018} to solve the above minimization problem.  
DA is a hardware that performs annealed  MCMC search to solve a combinatorial optimization problem with an objective function in the form of an Ising energy function.  
It has been observed that using DA results in significant speed up and improvement  compared to the state of the art simulated annealing solutions for fully connected spin glass problems \cite{aramon2018}.

\textcolor{black}{
DA is a massively parallel hardware architecture for solving combinatorial optimization problems and according to  \cite{aramon2018}, DA currently exhibits a time-to-solution speedup of roughly two orders of magnitude for fully connected spin-glass problems over the implementations of simulated annealing and parallel tempering Monte Carlo. Thus, the main complexity is in finding and formulating our problem in an Ising model format. 
}

%% file: Performance_evaluation.tex
\section{Performance Evaluation}
\label{sec.performance_evaluation}


In our simulations, the wireless channel is generated using the popular WIM2 simulator\cite{bultitude20074}, which is a powerful tool for simulating wireless channel.  In the WIM2 simulator, instead of simulating at link level, the whole scenario is simulated. This simulator gives several options to modify parameters such as simulation environment (indoor, outdoor, etc.), location of access points and users, etc. In addition, it can simulate the case where strong multipath interference is available. In our simulations, there are 16 multipath clusters (including LoS).

\begin{figure}
  \includegraphics[clip,width=\columnwidth]{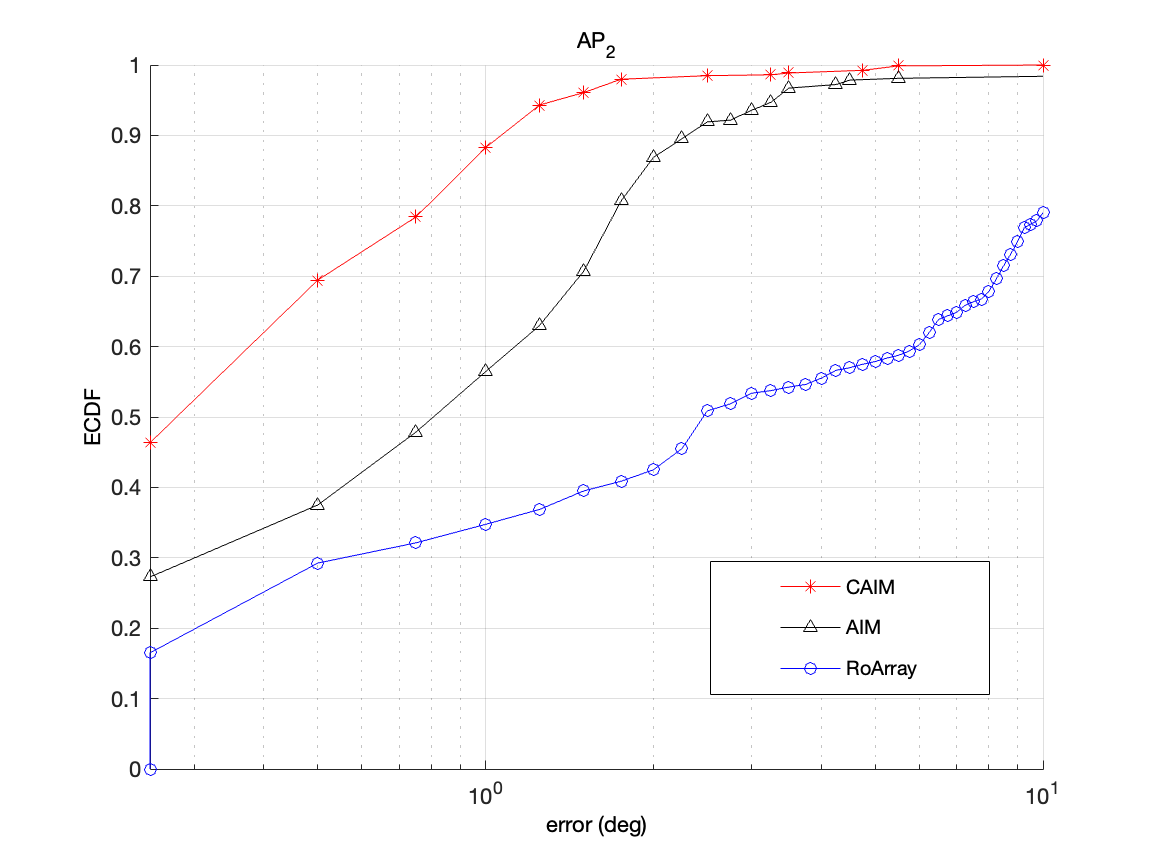}%
\caption{ECDF of AoA error for $AP_2$ for on-the-grid scenario.
}
\label{fig:fig1}
\end{figure}

\begin{table}
\centering
 \begin{tabular}{|c|c |c| c|c|c|} 
 \hline
  & $AP_1$ & $AP_2$ & $AP_3$& $AP_4$&$AP_5$ \\
 \hline
 CAIM & $0.25^{\circ}$ & $0.28^{\circ}$ & $0.27^{\circ}$& $0.28^{\circ}$ &$0.3^{\circ}$\\ 
 \hline
 AIM & $1.56^{\circ}$ & $0.81^{\circ}$& $0.5^{\circ}$& $0.62^{\circ}$ & $0.8^{\circ}$ \\
 \hline
 RoArray& $1.67^{\circ}$ & $1.98^{\circ}$ & $2.9^{\circ}$&$2.25^{\circ}$&$1.69^{\circ}$\\
 \hline
 \end{tabular}
  \caption{Median accuracy for all APs for on-the-grid scenario }
  \label{tab:table1}
\end{table}

To evaluate our proposed method, we generate the wireless channel with the WIM2 simulator for a target at far-field. We consider different scenarios with different access point rotations with respect to the global coordinate system (GCS). In each scenario, $\mbox{SNR} = 0$dB and the number of elements of the antenna arrays, $M$, is 8. The angular resolution of our search space is $0.25^{\circ}$. Therefore, $N_r = 720$. In our simulations, we consider $P=5$ access points in the environment. All experiments were conducted on the DA environment.

In Fig. \ref{fig:fig1}      , we represent the empirical cumulative distribution function (ECDF) of the AoA estimation error for one of the access points ($AP_2$). In this figure, the rotations of access points are $\{ \phi_1 = 120^{\circ}, \phi_2 = 225^{\circ}, \phi_3 = 200^{\circ}, \phi_4 = 150^{\circ}, \phi_5 = 230^{\circ} \}$. We consider RoArray \cite{Gong2019} and AIM \cite{Han2020} as our benchmarks. RoArray  formulates the AoA estimation problem as an $\textit{l}_1$ minimization problem and solves it for each of the $P$ access points. Then, it combines all the AoA estimates by forming a convex MMSE optimization problem.  AIM uses DA to solve the problem (\ref{back.eq5}) for each access point separately. The median accuracies for all the access points are demonstrated in Table~\ref{tab:table1}. The average median accuracy of CAIM is  $0.27^{\circ}$ while the average median accuracy for AIM and RoArray are $0.85^{\circ}$ and $2.1^{\circ}$, respectively.

\begin{figure}[t]
  \includegraphics[clip,width=\columnwidth]{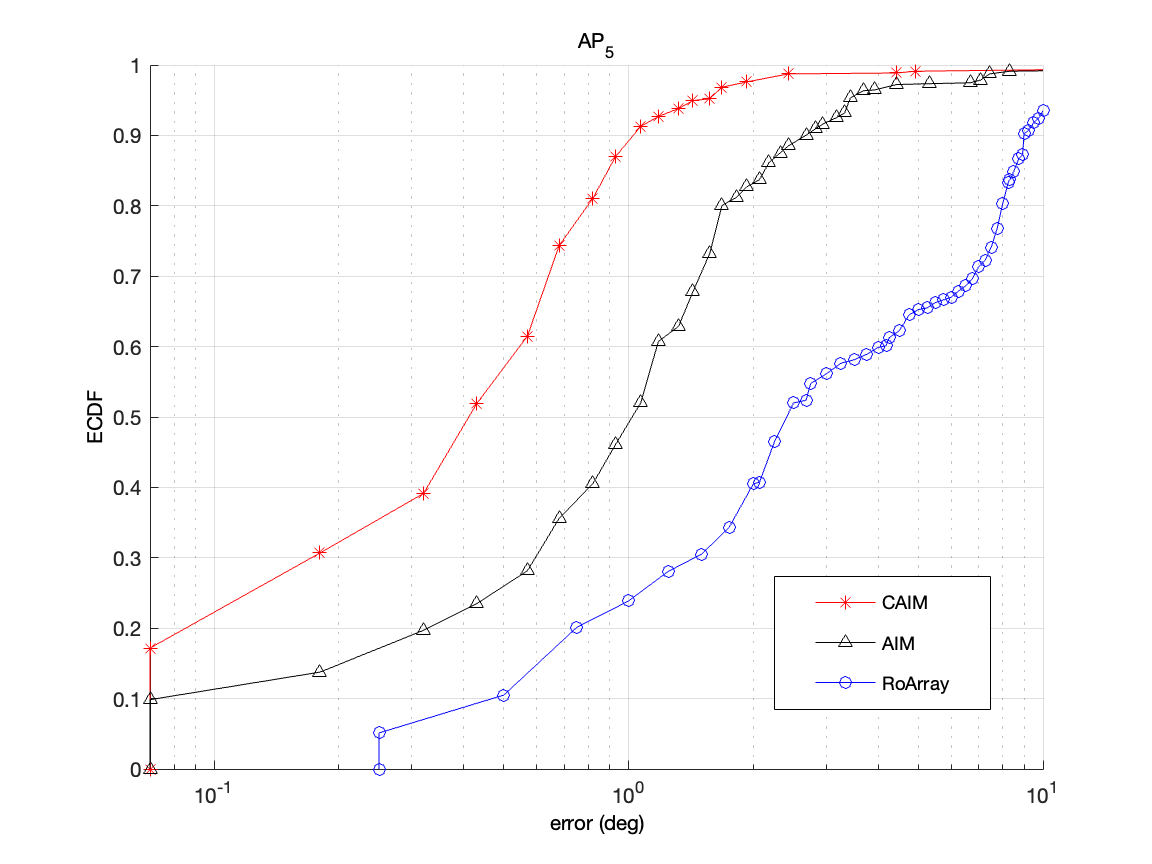}%
\caption{ECDF of AoA error for $AP_5$ for off-the-grid scenario.
}
\label{fig:fig2}
\end{figure}

In Fig.~\ref{fig:fig2}, the CDF of the AoA estimation error for an off-the-grid scenario for $AP_5$ is displayed. In this scenario, the rotations of access points are $\{ \phi_1 = 210.2^{\circ}, \phi_2 = 170.8^{\circ}, \phi_3 = 110.45^{\circ}, \phi_4 = 140.55^{\circ}, \phi_5 = 225.32^{\circ} \}$. Also, the median accuracy for all the access points is shown in Table~\ref{tab:table2}. For this scenario, the average median accuracy for our approach, AIM, and RoArray are $0.33^{\circ}$, $0.94^{\circ}$ and $2.23^{\circ}$, respectively. \textcolor{black}{In the off-the-grid case, all the estimations have at least an error because the ground truth is located between two adjacent grid points. However, the off-the-grid case can be the first order approximation of the true observation model. Therefore, the performance of our method does not degrade in comparison to other methods.}

\begin{table}
\centering
 \begin{tabular}{|c|c |c| c|c|c|} 
 \hline
  & $AP_1$ & $AP_2$ & $AP_3$& $AP_4$&$AP_5$ \\
 \hline
 CAIM & $0.32^{\circ}$ & $0.35^{\circ}$ & $0.33^{\circ}$& $0.3^{\circ}$ &$0.34^{\circ}$\\ 
 \hline
 AIM& $0.74^{\circ}$ & $0.4^{\circ}$ & $2^{\circ}$&$0.56^{\circ}$&$1.01^{\circ}$\\
 \hline
 RoArray & $4.25^{\circ}$ & $0.98^{\circ}$& $1.41^{\circ}$& $2.8^{\circ}$ & $1.71^{\circ}$ \\
 \hline
 \end{tabular}
  \caption{Median accuracy for all APs for off-the-grid scenario}
  \label{tab:table2}
\end{table}

\begin{figure}
    \centering
    \includegraphics[clip,width=\columnwidth]{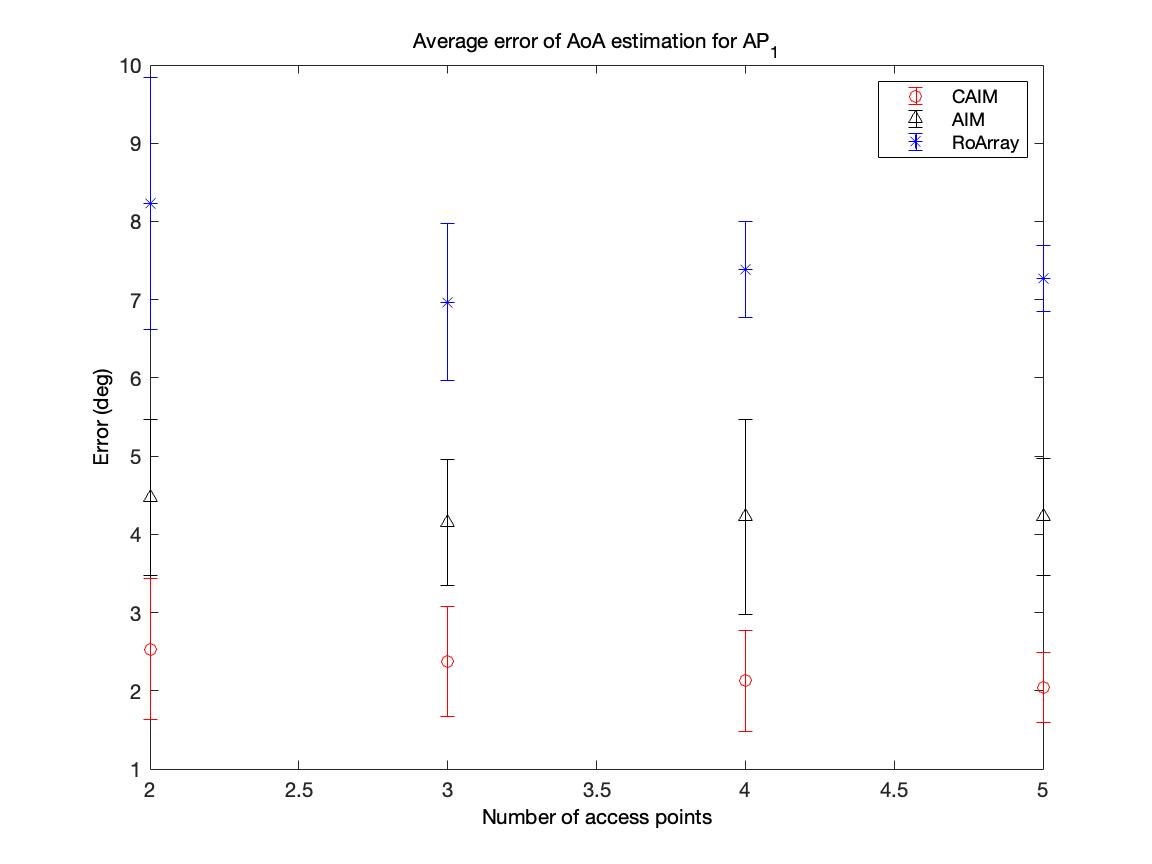}
    \caption{Comparison of the average error in degrees of CAIM with RoArray \cite{Gong2019} and AIM \cite{Han2020}.}
    \label{fig:fig3}
\end{figure}

In Fig.~\ref{fig:fig3}, the effect of increasing the number of access points on the average error of AoA estimation for one of the access points ($AP_1$) is investigated. As the number of access points increases, the average error of CAIM decreases. In addition, the average error of CAIM is less than that of AIM and RoArray.

%% file: Conclusion.tex
\section{Conclusion}
\label{sec.conclusion}
This paper  develops a method for cooperative AoA estimation using Ising energy model. In this paper, we propose a combination method to co-process the received signals from all the access points considering the angular relation of each pair of access poin ts. This process is formulated as an optimization problem with the help of Compressive Sensing. Then, the problem is transformed to an Ising model problem and is efficiently solved using Digital Annealer. Simulation results show that co-processing the received signals from multiple access points can  increase the resolution of AoA estimation significantly.